\documentstyle[12pt]{article}
\def\lbkt{\left[}\def\rbkt{\right]}
\def\beq{\begin{equation}}\def\eeq{\end{equation}}
\def\bea{\begin{eqnarray}}\def\eea{\end{eqnarray}}
\def\bean{\begin{eqnarray*}}\def\eean{\end{eqnarray*}}
\def\mb0{\makebox(0,0)}
\def\dsp{\displaystyle}

\newcommand{\dotl}[1]{\multiput(#1)(1.5,2){8}{\circle*{0.1}}}
\begin{document}
\begin{titlepage}
\hfill SNUTP 93-09 \vskip 1in
\begin{center}\begin{tabular}{c}
{\bf\LARGE Honeycomb Lattice Solvable Models}\\ \\ \\
Kyung-Hoon Kwon and Doochul Kim\\ \\
{\it Department of Physics and Center for Theoretical Physics}\\
{\it Seoul National University, Seoul 151-742, Korea}\\
\end{tabular}\end{center} \vskip 1in
\begin{abstract}
We construct solvable models on the honeycomb lattice by combining
three faces of the square lattice solvable models into a hexagon face. These
models contain two independent, anisotropy controlling, spectral parameters
and their transfer matrices with different spectral parameters commute with
each other.
At the critical point, the finite-size scaling of the transfer
matrix spectrum for the honeycomb lattice models is written in
terms of the quantities obtained from the finite-size scaling
of the square lattice solvable models.
We study in detail the phase transition properties of two models
based on the interacting hard square model and the magnetic hard square
model, respectively.
The models, in general, can be extended to the IRF version of the
Z-invariant models of Baxter.
\end{abstract}
\end{titlepage}

\renewcommand{\thesection}{\Roman{section}}

\section{Introduction}

\def\pf{\frac{\pi}{5}}
\def\dpf{\frac{2\pi}{5}}

The Yang-Baxter equation, also called the star-triangle relation (STR)
has been one of the most useful tools for analytical
calculation in the statistical lattice models\cite{Baxter}.
A lattice model is called solvable if its Boltzmann weights
satisfy the STR.
There are a variety of solvable models.
These models are all defined on the square lattice.
In this work, we construct solvable models on the honeycomb
lattice by combining three faces of the square lattice solvable models
into a hexagon.
The working is best illustrated from the perspective
of the kagome lattice eight-vertex model and its
Ising spin version on its dual lattice as follows.

In 1978, Baxter suggested the Z-invariant model on an arbitrary
planar lattice where any three lines do not intersect
at a point.\cite{Royal} The intersections of the lines
are regarded as lattice sites
and an arrow is assigned to each section of line divided by two other
lines.
A model is said to be Z-invariant if its interaction satisfies
the STR of vertex model for
any triangle of the planar lattice.
Here Z-invariant implies that the partition function
$\cal Z$ remains unchanged after sliding a line over a vertex point.
One of the examples of Z-invariant model
is the kagome lattice eight-vertex model
which is investigated in Ref.\cite{Baxter}.

On the kagome lattice, there exist three types of vertices.
For solvable eight-vertex model on the kagome lattice,
each type of vertices is described by different spectral parameters
and  is related to other types by the STR.
The total free energy  of this model is
simple sum of three free energies each of which are
obtained from the regular square lattice eight-vertex
model of the corresponding spectral parameter.
The two-point correlation function $\langle\sigma_r\sigma_m\rangle$
of two spins $\sigma_r$ and $\sigma_m$ located at site $r$ and $m$
depends only on one spectral parameter, if the sites $r$ and $m$
lie between two adjacent parrallel lines and
the system is large enough.

The kagome lattice eight-vertex model can be trans\-form\-ed
to the inter\-action-round-a-face (IRF) model on its dual lattice.
The dual lattice sites can be divided into two kinds of sites;
one on the face of the hexagons and the up triangles and
the other on the face of the down triangles of
the kagome lattice.
The former forms the honeycomb sublattice and the
latter the triangular sublattice.
The dual lattice of the kagome lattice will be
called the honeycomb-triangular lattice.
We can assign Ising spins  to the honeycomb-triangular lattice points
according to the direction of arrows of the original vertex model.
Since the vertex weight of the eight-vertex model is
determined by four arrows surrounding a lattice point,
the corresponding Ising model
possesses both two and four spin interactions.
When one takes the partial trace over the spins
on the triangular sublattice, the resulting system is
an Ising model on the remaining honeycomb lattice
with complicated interactions.
In this way, one can obtain the honeycomb lattice Ising model
in which a hexagon is the unit of the IRF model.

We write the face weight of the Ising model
corresponding to a vertex with spectral parameter $u$ as $w(a,b,c,d|u)$
where $a$, $b$, $c$ and $d$ denote the spin states
located at lower-left, lower-right, upper-right, and upper-left corners
of a face, respectively,
and denote by $W_{\rm h}(a,b,c;d,e,f)$
the hexagonal face weight for a face whose corners have
the spin values, $a$, $b$, $c$, $d$, $e$ and $f$, counterclockwisely.
Then the face weight for a hexagonal face is written
in terms of the three spectral parameters $u$, $v$ and $z$
as (See Section II and Fig.\ref{hexagon}.)
\beq
W_{\rm h}(a,b,c;d,e,f)=\sum_g w(e,g,c,d|u)w(g,a,b,c|v)w(f,a,g,e|z)
\; .
\label{def}
\eeq
The summation runs over the possible spin values $\pm 1$.
Following the condition for the Z-invariance\cite{Baxter},
the spectral parameter $z$ is
connected to the other parameters $u$ and $v$ by $z=u-v$.
In the notation of $W_{\rm h}$,
two independent variables $u$ and $v$
will be specified as $W_{\rm h}(a,b,c;d,e,f|u,v)$.

An important observation is that the model obtained in this
way is also solvable as we discuss in Section II.
In fact, one can construct honeycomb lattice solvable models (HLSM)
in this way starting from arbitrary weights $w(a,b,c,d|u)$
of solvable square lattice models.
The purpose of this paper is to formulate generally the
HLSM and study their properties.
This paper is organized as follows.
In Section II, we define the HLSM in a general setting and
show the commutivity of the row-to-row transfer matrix.
We also derive the general form for the bulk and the
finite-size corrections of the transfer matrix spectra.
In Section III and IV, we study the HLSM's based on the
interacting hard square and the magnetic hard square, respectively.
The ground state symmetries are mapped in the $(u$-$v)$ plane
and, in the case of the magnetic hard squares, the transfer
matrix spectrum is analyzed.
In Section V, we extend the HLSM to a hierarchy of models which have
larger unit cells.
We conclude with a discussion in Section VI.

\section{Honeycomb lattice solvable models}

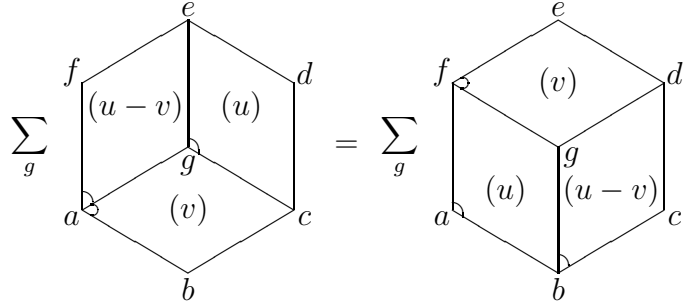
\begin{figure}
\begin{center}
\begin{picture}(250,150)(0,50)
\put(0,142){\mb0{$\displaystyle{\sum_g}$}}
\multiput(60,190)(140,0){2}{\line(-5,-3){40}}
\multiput(60,190)(140,0){2}{\line(5,-3){40}}
\multiput(20,118)(140,0){2}{\line(0,1){48}}
\multiput(20,118)(140,0){2}{\line(5,-3){40}}
\multiput(100,118)(140,0){2}{\line(0,1){48}}
\multiput(100,118)(140,0){2}{\line(-5,-3){40}}
\multiput(16,115)(140,0){2}{\mb0{$a$}}
\multiput(60,88)(140,0){2}{\mb0{$b$}}
\multiput(104,115)(140,0){2}{\mb0{$c$}}
\multiput(104,170)(140,0){2}{\mb0{$d$}}
\multiput(60,194)(140,0){2}{\mb0{$e$}}
\multiput(16,170)(140,0){2}{\mb0{$f$}}
\put(60,190){\line(0,-1){48}}
\put(60,135){\mb0{$g$}}
\put(20,118){\line(5,3){40}}
\put(100,118){\line(-5,3){40}}
\put(40,157){\mb0{$(u-v)$}}
\put(80,157){\mb0{$(u)$}}
\put(60,118){\mb0{$(v)$}}
\put(60,139){\oval(8,12)[tr]}
\put(20,121){\oval(8,8)[tr]}
\put(23,118){\oval(6,4)[r]}
\put(120,142){\mb0{$=$}}
\put(140,142){\mb0{$\displaystyle{\sum_g}$}}
\put(200,142){\line(0,-1){48}}
\put(200,142){\line(5,3){40}}
\put(200,142){\line(-5,3){40}}
\put(200,166){\mb0{$(v)$}}
\put(163,166){\oval(6,4)[r]}
\put(180,125){\mb0{$(u)$}}
\put(220,125){\mb0{$(u-v)$}}
\put(205,138){\mb0{$g$}}
\put(160,115){\oval(8,12)[tr]}
\put(200,97){\oval(8,8)[tr]}
\end{picture}
\end{center}
\caption{Graphical representation of the star-triangle relation.}
\label{str}
\end{figure}

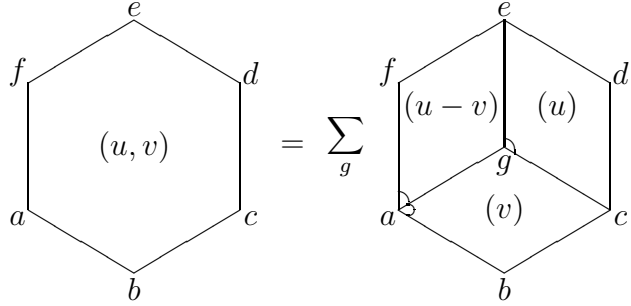
\begin{figure}
\begin{center}
\begin{picture}(250,120)(0,80)
\multiput(60,190)(140,0){2}{\line(-5,-3){40}}
\multiput(60,190)(140,0){2}{\line(5,-3){40}}
\multiput(20,118)(140,0){2}{\line(0,1){48}}
\multiput(20,118)(140,0){2}{\line(5,-3){40}}
\multiput(100,118)(140,0){2}{\line(0,1){48}}
\multiput(100,118)(140,0){2}{\line(-5,-3){40}}
\put(60,142){\mb0{$(u,v)$}}
\multiput(16,115)(140,0){2}{\mb0{$a$}}
\multiput(60,88)(140,0){2}{\mb0{$b$}}
\multiput(104,115)(140,0){2}{\mb0{$c$}}
\multiput(104,170)(140,0){2}{\mb0{$d$}}
\multiput(60,194)(140,0){2}{\mb0{$e$}}
\multiput(16,170)(140,0){2}{\mb0{$f$}}
\put(120,142){\mb0{$=$}}
\put(140,142){\mb0{${\displaystyle \sum_g}$}}
\put(200,190){\line(0,-1){48}}
\put(200,135){\mb0{$g$}}
\put(160,118){\line(5,3){40}}
\put(240,118){\line(-5,3){40}}
\put(180,157){\mb0{$(u-v)$}}
\put(220,157){\mb0{$(u)$}}
\put(200,118){\mb0{$(v)$}}
\put(200,139){\oval(8,12)[tr]}
\put(160,121){\oval(8,8)[tr]}
\put(163,118){\oval(6,4)[r]}
\end{picture}
\end{center}
\caption{Definition of the Boltzmann face weights
in the honeycomb lattice solvable model.}
\label{hexagon}
\end{figure}

We start with the definition of the
honeycomb lattice solvable models (HLSM)
which are the IRF models constructed by using solvable
models on the square lattice.
Let ($a,b,c,d$) denotes a spin configuration of a unit square
for a lattice spin model on the square lattice.
Values of $a$, $b$, $c$, $d$ range over the
set of spin states depending on models.
The face weights of solvable IRF models are denoted
as $w(a,b,c,d|u)$ where $u$ is the spectral parameter and its
dependences on other interaction parameters are implicit.
$w$ is the solution of the STR which is shown graphically in
Fig.\ref{str},
and satisfy the standard reflection symmetries
\[w(a,b,c,d|u)=w(c,b,a,d|u)=w(a,d,c,b|u)\]
and the initial condition
\[w(a,b,c,d|0)=\delta_{ac}\;.\]
For any solution $w(a,b,c,d|u)$ obtained from the STR,
we can construct the HLSM whose
hexagonal face weights
are given by Eq.(\ref{def}).
This definition of the hexagonal face weights is shown diagramatically
in Fig.\ref{hexagon}.
Since the square lattice face weights satisfy the STR, we get
the identity of
\beq
\begin{array}{l}
\dsp{W_{\rm h}(a,b,c;d,e,f|u,v)} \\
\hskip 1.5cm\dsp{=\sum_gw(f,a,b,g|u)w(g,b,c,d|u-v)w(e,f,g,d|v)\;.}
\end{array}
\label{sym}
\eeq
Using the diagonal reflection symmetry to Eq.(\ref{sym}),
we have the symmetry relation
\[ W_{\rm h}(a,b,c;d,e,f|u,v)=W_{\rm h}(d,e,f;a,b,c|u,v)\;.\]

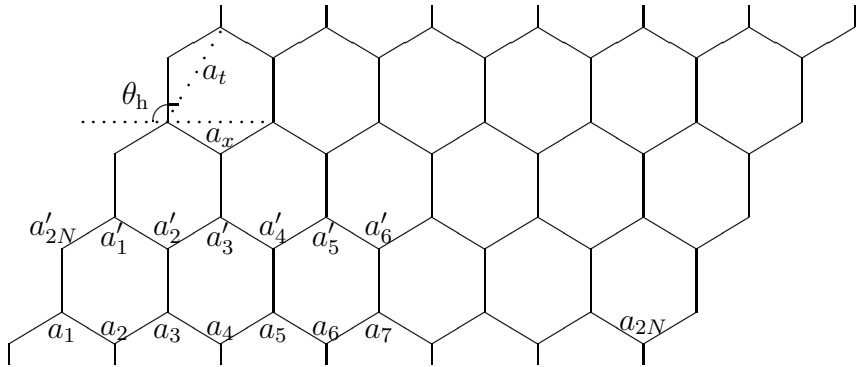
\begin{figure}
\begin{center}
\begin{picture}(250,140)(-50,60)
\multiput(40,190)(40,0){7}{\line(-5,-3){20}}
\multiput(40,190)(40,0){6}{\line(5,-3){20}}
\multiput(20,154)(40,0){7}{\line(0,1){24}}
\multiput(20,154)(40,0){6}{\line(5,-3){20}}
\multiput(20,154)(40,0){7}{\line(-5,-3){20}}
\multiput(-12,154)(4,0){18}{\circle*{0.3}}
\multiput(22,158)(3,5){7}{\circle*{0.3}}
\put(24,154){\oval(19,13)[tl]}
\put(8,164){\mb0{$\theta_{\rm h}$}}
\put(40,148){\mb0{$a_x$}}
\put(38,172){\mb0{$a_t$}}
\multiput(0,118)(40,0){7}{\line(0,1){24}}
\put(0,110){\mb0{$a^\prime_1$}}
\put(40,110){\mb0{$a^\prime_3$}}
\put(80,110){\mb0{$a^\prime_5$}}
\put(-23,114){\mb0{$a^\prime_{2N}$}}
\put(20,114){\mb0{$a^\prime_2$}}
\put(60,114){\mb0{$a^\prime_4$}}
\put(100,114){\mb0{$a^\prime_6$}}
\put(-20,74){\mb0{$a_1$}}
\put(20,74){\mb0{$a_3$}}
\put(60,74){\mb0{$a_5$}}
\put(100,74){\mb0{$a_7$}}
\put(0,75){\mb0{$a_2$}}
\put(40,75){\mb0{$a_4$}}
\put(80,75){\mb0{$a_6$}}
\put(200,77){\mb0{$a_{2N}$}}
\multiput(0,118)(40,0){6}{\line(5,-3){20}}
\multiput(0,118)(40,0){7}{\line(-5,-3){20}}
\multiput(-20,82)(40,0){7}{\line(0,1){24}}
\multiput(-20,82)(40,0){6}{\line(5,-3){20}}
\multiput(-20,82)(40,0){7}{\line(-5,-3){20}}
\multiput(-40,70)(40,0){7}{\line(0,-1){8}}
\multiput(40,190)(40,0){7}{\line(0,1){8}}
\end{picture}
\end{center}
\caption{Distribution of the spins to the honeycomb lattice sites.
$a_x$, $a_t$ and $\theta_{\rm h}$ are the parameters appearing
in the finite-size scaling and not to be confused with the
spin state labelings.}
\label{transfer}
\end{figure}

We consider a honeycomb lattice system with $M$ rows
each of which is made of $N$ hexagons as shown in Fig.\ref{transfer}.
To the lower and the upper sites of a row,
we assign the spin configurations
$\vec a=(a_1,$ $a_2,$ $\cdots,$ $a_{2N})$
and $\vec a^\prime=(a_1^\prime,$ $a_2^\prime,$
$\cdots,$ $a_{2N}^\prime)$, respectively.
The $a_i$'s and $a^\prime_i$'s are the possible spin states.
For convenience, we count the sites so that the set of spins
$\{a_3,$ $a_4,$ $a_5,$ $a_4^\prime,$
$a_3^\prime,$ $a_2^\prime\}$ constitutes a hexagonal face.
In this paper, we restrict ourselves to
the periodic boundary condition; $a_{2N+1}=a_1$
and $a_{2N+1}^\prime=a_1^\prime$.

Defining the transfer matrix ${\bf V}_{\vec a\vec a^\prime}(u,v)$
as
\[{\bf V}_{\vec a\vec a^\prime}(u,v)=\prod_{\mbox{odd}\ i<2N}
W_{\rm h}(a_i, a_{i+1}, a_{i+2}; a_{i+1}^\prime,
a_i^\prime, a_{i-1}^\prime|u,v)\;,\]
the partition function $\cal Z$ of the system with $M$ rows
is written in terms of the transfer matrix ${\bf V}(u,v)$ as
\[{\cal Z}=\mbox{tr} [{\bf V}(u,v)]^M\;.\]
The transfer matrix {\bf V} of the honeycomb lattice model
is the function of two independent spectral variables $u$ and $v$.
Using the STR, the commuting relation of
\[[{\bf V}(u,v),{\bf V}(u^\prime,v)]=0\]
can be proved for any values of $u$ and $u^\prime$.
This is best illustrated graphically as shown in Fig.\ref{com1}.
\begin{figure}
\begin{center}
\begin{picture}(250,190)(-50,0)
\multiput(80,190)(80,0){2}{\line(-5,-3){40}}
\multiput(0,190)(80,0){3}{\line(5,-3){40}}
\multiput(0,190)(80,0){3}{\line(0,-1){48}}
\multiput(0,139)(80,0){3}{\oval(8,12)[tr]}
\multiput(40,67)(80,0){2}{\oval(8,12)[tr]}
\multiput(20,150)(80,0){3}{\mb0{$u$}}
\multiput(60,150)(80,0){2}{\mb0{$u-v$}}
\multiput(0,118)(80,0){3}{\mb0{$v$}}
\multiput(40,118)(80,0){3}{\line(0,1){48}}
\multiput(40,118)(80,0){2}{\line(5,-3){40}}
\multiput(40,121)(80,0){2}{\oval(8,8)[tr]}
\multiput(0,49)(80,0){3}{\oval(8,8)[tr]}
\multiput(43,118)(80,0){2}{\oval(6,4)[r]}
\multiput(3,46)(80,0){3}{\oval(6,4)[r]}
\multiput(40,118)(80,0){3}{\line(0,-1){48}}
\multiput(40,118)(80,0){2}{\line(5,3){40}}
\multiput(40,118)(80,0){3}{\line(-5,-3){40}}
\multiput(40,118)(80,0){3}{\line(-5,3){40}}
\multiput(20,78)(80,0){3}{\mb0{$u^\prime-v$}}
\multiput(60,78)(80,0){2}{\mb0{$u^\prime$}}
\multiput(40,46)(80,0){3}{\mb0{$v$}}
\multiput(0,46)(80,0){3}{\line(0,1){48}}
\multiput(0,46)(80,0){3}{\line(5,-3){40}}
\multiput(0,46)(80,0){3}{\line(5,3){40}}
\multiput(80,46)(80,0){2}{\line(-5,3){40}}
\multiput(80,46)(80,0){2}{\line(-5,-3){40}}
\put(100,10){\mb0{(a)}}
\end{picture}
\begin{picture}(250,200)(-50,0)
\multiput(80,190)(80,0){2}{\line(-5,-3){40}}
\multiput(0,190)(80,0){3}{\line(5,-3){40}}
\multiput(0,142)(80,0){3}{\line(0,-1){48}}
\multiput(0,142)(80,0){3}{\line(5,3){40}}
\multiput(80,142)(80,0){2}{\line(-5,3){40}}
\multiput(40,67)(80,0){2}{\oval(8,12)[tr]}
\multiput(40,115)(80,0){2}{\oval(8,12)[tr]}
\multiput(20,130)(80,0){3}{\mb0{$u-v$}}
\multiput(60,130)(80,0){2}{\mb0{$u$}}
\multiput(0,166)(80,0){3}{\mb0{$v$}}
\multiput(40,118)(80,0){3}{\line(0,1){48}}
\multiput(40,118)(80,0){2}{\line(5,-3){40}}
\multiput(0,49)(80,0){3}{\oval(8,8)[tr]}
\multiput(0,97)(80,0){3}{\oval(8,8)[tr]}
\multiput(3,46)(80,0){3}{\oval(6,4)[r]}
\multiput(43,166)(80,0){2}{\oval(6,4)[r]}
\multiput(40,118)(80,0){3}{\line(0,-1){48}}
\multiput(40,118)(80,0){3}{\line(-5,-3){40}}
\multiput(20,78)(80,0){3}{\mb0{$u^\prime-v$}}
\multiput(60,78)(80,0){2}{\mb0{$u^\prime$}}
\multiput(40,46)(80,0){3}{\mb0{$v$}}
\multiput(0,46)(80,0){3}{\line(0,1){48}}
\multiput(0,46)(80,0){3}{\line(5,-3){40}}
\multiput(0,46)(80,0){3}{\line(5,3){40}}
\multiput(80,46)(80,0){2}{\line(-5,3){40}}
\multiput(80,46)(80,0){2}{\line(-5,-3){40}}
\put(100,10){\mb0{(b)}}
\end{picture}
\end{center}
\caption{Proof of the relation $[{\bf V}(u,v),{\bf V}(u^\prime,v)]=0$.
(a) Two rows with the transfer matrix
${\bf V}(u,v)$ and ${\bf V}(u^\prime,v)$, respectively.
(b) By the STR, we have moved the faces with the spectral
parameter $v$ over the faces with the parameter $u$ and $u-v$.
(c) Two rows in the middle
have been interchanged, using the properties of square lattice
solvable models. (d) By the STR, the uppermost faces with the
spectral parameter $v$ go down beyond the faces with the
spectral parameter $u^\prime$ and $u^\prime-v$.}
\label{com1}
\end{figure}
\begin{figure}\begin{center}
\begin{picture}(250,190)(-50,0)
\multiput(80,190)(80,0){2}{\line(-5,-3){40}}
\multiput(0,190)(80,0){3}{\line(5,-3){40}}
\multiput(0,142)(80,0){3}{\line(0,-1){48}}
\multiput(0,142)(80,0){3}{\line(5,3){40}}
\multiput(80,142)(80,0){2}{\line(-5,3){40}}
\multiput(40,67)(80,0){2}{\oval(8,12)[tr]}
\multiput(40,115)(80,0){2}{\oval(8,12)[tr]}
\multiput(20,130)(80,0){3}{\mb0{$u^\prime-v$}}
\multiput(60,130)(80,0){2}{\mb0{$u^\prime$}}
\multiput(0,166)(80,0){3}{\mb0{$v$}}
\multiput(40,118)(80,0){3}{\line(0,1){48}}
\multiput(40,118)(80,0){2}{\line(5,-3){40}}
\multiput(0,49)(80,0){3}{\oval(8,8)[tr]}
\multiput(0,97)(80,0){3}{\oval(8,8)[tr]}
\multiput(3,46)(80,0){3}{\oval(6,4)[r]}
\multiput(43,166)(80,0){2}{\oval(6,4)[r]}
\multiput(40,118)(80,0){3}{\line(0,-1){48}}
\multiput(40,118)(80,0){3}{\line(-5,-3){40}}
\multiput(20,78)(80,0){3}{\mb0{$u-v$}}
\multiput(60,78)(80,0){2}{\mb0{$u$}}
\multiput(40,46)(80,0){3}{\mb0{$v$}}
\multiput(0,46)(80,0){3}{\line(0,1){48}}
\multiput(0,46)(80,0){3}{\line(5,-3){40}}
\multiput(0,46)(80,0){3}{\line(5,3){40}}
\multiput(80,46)(80,0){2}{\line(-5,3){40}}
\multiput(80,46)(80,0){2}{\line(-5,-3){40}}
\put(100,10){\mb0{(c)}}
\end{picture}
\begin{picture}(250,200)(-50,0)
\multiput(80,190)(80,0){2}{\line(-5,-3){40}}
\multiput(0,190)(80,0){3}{\line(5,-3){40}}
\multiput(0,190)(80,0){3}{\line(0,-1){48}}
\multiput(0,139)(80,0){3}{\oval(8,12)[tr]}
\multiput(40,67)(80,0){2}{\oval(8,12)[tr]}
\multiput(20,150)(80,0){3}{\mb0{$u^\prime$}}
\multiput(60,150)(80,0){2}{\mb0{$u^\prime-v$}}
\multiput(0,118)(80,0){3}{\mb0{$v$}}
\multiput(40,118)(80,0){3}{\line(0,1){48}}
\multiput(40,118)(80,0){2}{\line(5,-3){40}}
\multiput(40,121)(80,0){2}{\oval(8,8)[tr]}
\multiput(0,49)(80,0){3}{\oval(8,8)[tr]}
\multiput(43,118)(80,0){2}{\oval(6,4)[r]}
\multiput(3,46)(80,0){3}{\oval(6,4)[r]}
\multiput(40,118)(80,0){3}{\line(0,-1){48}}
\multiput(40,118)(80,0){2}{\line(5,3){40}}
\multiput(40,118)(80,0){3}{\line(-5,-3){40}}
\multiput(40,118)(80,0){3}{\line(-5,3){40}}
\multiput(20,78)(80,0){3}{\mb0{$u-v$}}
\multiput(60,78)(80,0){2}{\mb0{$u$}}
\multiput(40,46)(80,0){3}{\mb0{$v$}}
\multiput(0,46)(80,0){3}{\line(0,1){48}}
\multiput(0,46)(80,0){3}{\line(5,-3){40}}
\multiput(0,46)(80,0){3}{\line(5,3){40}}
\multiput(80,46)(80,0){2}{\line(-5,3){40}}
\multiput(80,46)(80,0){2}{\line(-5,-3){40}}
\put(100,10){\mb0{(d)}}
\end{picture}
\end{center}
\end{figure}
\setcounter{figure}{4}
In a similar manner, we also get the commuting relation of
\[[{\bf V}(u,v),{\bf V}(u,v^\prime)]=0\]
for different $v$ and $v^\prime$.
Those relations imply the integrability of the model.

The free energy $f_{\rm h}$ per a hexagon face is given by
\[f_{\rm h}(u,v)=f(u)+f(v)+f(u-v)\;,\]
where $f(u)$ is the free energy per square of the square
lattice system defined by the original face weight $w(a,b,c,d|u)$.
This relation is derived
in the same way as how the free energy of the kagome lattice
eight-vertex model is obtained in \cite{Baxter}.

Assuming the honeycomb lattice model has the conformal symmetry
at the critical point, the
finite-size scaling of the transfer matrix spectrum can be predicted by
the conformal field theory.
For the $\alpha$-th eigenvalue $\Lambda_\alpha$ of the
transfer matrix ${\bf V}(u,v)$, we expect the
finite-size scaling form as follows\cite{KimPearce},
\beq
- \ln |\Lambda_\alpha|=Nf_{\rm h}+\frac{2\pi}{N}(x_\alpha-\frac{c}{12})
\frac{a_t}{a_x}\sin\theta_{\rm h}
\label{fss}
\eeq
under the periodic boundary condition.
Here $f_{\rm h}$ is the bulk free energy,
$c$ is the central charge and $x_\alpha$ is the scaling dimension
associated with the corresponding correlation function.
$a_x$ is the width of one hexagon in the horizontal direction
and $a_t$ is the length in the time direction.
$a_x$ and $a_t$ are illustrated in Fig.\ref{transfer}.
$\theta_{\rm h}$ is the angle depending on the anisotropy
of the interaction\cite{KimPearce}.
For general $u$ and $v$, the lattice should be deformed to make
the system isotropic in the continuum limit  so that
$a_t$, $a_x$ and $\theta_{\rm h}$ take non-trivial values.

In Ref.\cite{KimPearce}, it was shown that each unit square of
solvable square lattice models is deformed in such
a way that $a_t=a_x$ due to the diagonal reflection symmetry and
the deformation is characterized by the anisotropy angle $\theta$
which depends linearly on the spectral parameter.
Applying this result to each types of squares forming a hexagon,
one finds that the anisotropy parameters in Eq.(\ref{fss})
for the HLSM are given by
\bean
a_x&=&2a\cos\frac{\theta(v)}{2}\\
a_t&=&2a\cos\frac{\theta(u-v)}{2}\\
\theta_{\rm h}&=&\frac{\theta(u)}{2}\; ,
\eean
where $\theta(u)$ is the anisotropy angle
of the square lattice solvable model with the spectral
parameter $u$ and
$a$ is the lattice constant of a hexagon, that is,
the  distance between nearest neighbors of the honeycomb lattice.
These relations yield the finite-size scaling of the honeycomb lattice
model, in terms of the quantities obtained
in the square lattice solvable model as
\bea
- \ln|\Lambda_\alpha|&=&N\{f(u)+f(v)+f(u-v)\} \nonumber \\
& &
+\frac{2\pi}{N}(x_\alpha-\frac{c}{12})\frac{\cos\{\theta(u-v)/2\}}
{\cos\{\theta(v)/2\}}\sin\{\theta(u)/2\}\;. \label{fssH}
\eea

Due to the initial condition $w(a,b,c,d|0)=\delta_{ac}$,
the limit of $v\rightarrow 0$
reduces the honeycomb lattice to the square lattice system
of the spectral parameter $u$ with $2N$ faces in a row.
In the limit $v=0$, the above finite-size scaling
relation Eq.(\ref{fssH}) reduces to that of the square lattice model
with the parameter $u$ as it should be.

One of the simplest examples for the honeycomb lattice solvable models
is the standard Ising model.
The solvable Ising model with diagonal interaction on the square
lattice is defined by the interactions $K_u$ and $L_u$ through the face weight
\[ w(a,b,c,d|u)=\exp(K_uac+L_ubd)\; ,\]
where the $u$-dependences of $K_u$ and $L_u$ are given in the
standard references, e.g. Ref.\cite{Baxter}.
For the independent parameters $u$ and $v$,
the hexagonal weight is defined as
\bean
\lefteqn{W_{\rm h}(a,b,c;d,e,f|u,v)}& & \\
&=&\sum_g\exp(K_uec+K_{u-v}fg+K_vgb+L_ugd+L_{u-v}ea+L_vac)\; .
\eean
The STR simplifies this weight to
\bean
W_{\rm h}(a,b,c;d,e,f|u,v) &=&\exp(K_uce+L_{u-v}ea+L_vac) \\
& &\times\exp(K_ufb+L_{u-v}bd+L_vdf)
\eean
which shows that only the two body interactions between
next nearest neighbors appear on the honeycomb lattice.
Since there is no interaction between nearest neighbors,
the honeycomb lattice Ising model is equivalent
to two independent Ising models
on the triangular lattice.
However, for other solvable models, such simple decomposition
does not arise and following two sections are devoted to specific
models, i.e., the interacting hard square model and the magnetic
hard square model.

\section{Interacting hard square (IHS) model}

In this section, we consider the HLSM based on
the interacting hard square (IHS) model
which is a diagonally interacting lattice gas model where a
site cannot be occupied unless its nearest neighbors are all empty.
Each site has the value 0(empty) or 1(occupied) and the face
weight is described by the fugacity $z$, the interaction $L$ and $M$:
\bean
\lefteqn{w(a,b,c,d)}& & \\
&=&\left\{
\begin{array}{ll}
mz^{(a+b+c+d)/4}\exp(Lac+Mbd)t^{-a+b-c+d}&
\mbox{, if $ab=bc=cd=da=0$,}\\
0 &\mbox{, otherwise.}
\end{array}\right.
\eean
Here $m$ is a constant which can be adjusted for a suitable normalization.
The quantity $t$ is necessary so as to discover the solution satisfying
the STR. In the homogeneous square lattice where all the square faces
have the same interaction strength,
$t$ does not appear in the partition function.
However, on the honeycomb lattice, it can play a non-trivial role.
Baxter\cite{ihs} found the solutions to the STR on a special
manifold in the parameter space $(z,L,M,t)$.
The manifold is parametrized by the temperature-like variable $q$
and the spectral parameter $u$.
Interaction parameters, hence, the face weights are expressed by
elliptic functions with argument $u$ and nome $q^2$.
The face weights are defined when the spectral parameter, $u$
varies in the range of
$-\pi/5<u<2\pi/5$.

On the honeycomb lattice we will label six sites of a unit cell,
counterclockwisely from the uppermost corner, by sequential numbers
1 to 6.
For instance, in Fig.\ref{hexagon} we assign the site of the spin state $e$
as 1, $f$ as 2, and so on.
And we denote by $L_{ij}$ the interaction between the sites $i$ and
$j$.
$L_{ijk}$ is the three-spin interaction among the sites $i$, $j$ and
$k$.
Then the face weights for the honeycomb lattice IHS model
are given as
\bean
\lefteqn{W_{\rm h}(a,b,c;d,e,f|u,v)}& & \\
&=& m_0z_{\rm h}^{(a+b+c+d+e+f)/3}\exp(L_{14}eb+L_{25}fc+L_{36}ad) \\
& &\times\exp(L_{13}ea+L_{24}fb+L_{35}ac+L_{46}bd+L_{51}ce+L_{62}df)\\
& &\times\exp(L_{135}ace+L_{246}bdf)
\eean
where $z_{\rm h}$ is the fugacity and $m_0$
is the weight for the vacuum state, i.e. the state where
all six sites of a hexagon are vacant.
Actually the vacuum state of this honeycomb lattice model
contains the state with one particle at the center of the cell.
For the solvable IHS model, there is no four-spin interaction
on the honeycomb lattice.
{}From the exact solution of the square lattice,
we obtain the following functional form for the interactions:
\bean
m_0&=&1+\frac{\theta_1^2(\dpf)\theta_1(\pf-u)
\theta_1(u-v)\theta_1(v)}{\theta_1^2(\pf)
\theta_1(\dpf+u)\theta_1(\dpf+u-v)\theta_1(\dpf+v)}\\
z_{\rm h}&=&\frac{\theta_1(\dpf)\theta_1^2(u)
\theta_1^2(\pf-u+v)\theta_1^2(\pf-v)}
{\theta_1(\pf) \theta_1^2(\dpf+u)
\theta_1^2(\dpf+u-v)
\theta_1^2(\dpf+v)}m_0^3 \\
L_{13}&=&L_{46}\\
&=&\ln\lbkt \frac{\theta_1(\pf)\theta_1(\dpf+u-v)
\theta_1(\pf+u-v)}{\theta_1(\dpf)\theta_1^2(\pf-u+v)}
m_0\rbkt \\
L_{24}&=&L_{15}\\
&=&\ln\lbkt \frac{\theta_1(\pf)\theta_1(\dpf+u)
\theta_1(\dpf-u)}{\theta_1(\dpf)\theta_1^2(u)}m_0\rbkt \\
L_{35}&=&L_{26}\\
&=&\ln\lbkt \frac{\theta_1(\pf)\theta_1(\dpf+u)
\theta_1(\pf+v)}{\theta_1(\dpf)\theta_1^2(\pf-v)}m_0\rbkt \\
L_{14}&=&\ln\lbkt\frac{\theta_1(\pf)\theta_1(v)
\theta_1(\dpf+u)\theta_1(\dpf+u-v)}
{\theta_1(\dpf)\theta_1(u)\theta_1(\pf-u+v)
\theta_1(\dpf+v)}m_0\rbkt\\
L_{25}&=&\ln\lbkt\frac{\theta_1(\pf)\theta_1(u-v)
\theta_1(\dpf+u)\theta_1(\dpf+v)}
{\theta_1(\dpf)\theta_1(u)\theta_1(\pf-v)
\theta_1(\dpf+u-v)}m_0\rbkt \\
L_{36}&=&\ln\lbkt\frac{\theta_1(\pf)\theta_1(\pf-u)
\theta_1(\dpf+u-v)\theta_1(\dpf+v)}
{\theta_1(\dpf)\theta_1(\pf-u+v)\theta_1(\pf-v)
\theta_1(\dpf+u)}m_0\rbkt \\
L_{135}&=&L_{246}=-\ln m_0
\eean
where $\theta_1$ is the standard elliptic theta function.
It is defined by the spectral parameter $u$ and
the temperature-like parameter $q^2$ as
\[\theta_1(u)=\sin u\prod_{n=1}^\infty(1-2q^{2n}\cos 2u+q^{4n})
(1-q^{2n})\;.\]
For simplicity, we shall omit $q^2$ in the notation of $\theta_1$.
It is to be noted that the interactions show the symmetry under
the rotation by angle $180^\circ$ which leads to the property of
$W_{\rm h}(a,b,c;d,e,f|u,v)=W_{\rm h}(d,e,f;a,b,c|u,v)$.

\begin{figure}
\begin{center}
\begin{picture}(230,170)(-85,-55)
\put(31,-52){\line(-3,5){93}}
\put(-80,0){\line(1,0){210}}
\put(135,0){\mb0{$u$}}
\put(-65,109){\mb0{$v$}}
\multiput(-19,-52)(-12,20){6}{\line(-3,5){6}}
\multiput(81,-52)(-12,20){8}{\line(-3,5){6}}
\multiput(131,-52)(-12,20){8}{\line(-3,5){6}}
\multiput(-50,-42)(20,0){10}{\line(1,0){10}}
\multiput(-80,42)(20,0){9}{\line(1,0){10}}
\multiput(-80,84)(20,0){8}{\line(1,0){10}}
\put(-25,42){\line(1,0){100}}
\put(75,42){\line(-3,-5){50}}
\put(25,10){\mb0{(D)}}
\put(-20,-13){\mb0{(A)}}
\put(70,-13){\mb0{(B)}}
\put(25,58){\mb0{(C)}}
\multiput(-15.5,37)(9,0){10}{\circle{1}}
\multiput(-11,29.5)(9,0){9}{\circle{1}}
\multiput(-6.5,22)(9,0){8}{\circle{1}}
\multiput(-2,14.5)(9,0){3}{\circle{1}}
\multiput(34,14.5)(9,0){3}{\circle{1}}
\multiput(2.5,7)(9,0){2}{\circle{1}}
\multiput(38.5,7)(9,0){2}{\circle{1}}
\multiput(7,-0.5)(9,0){5}{\circle{1}}
\multiput(11.5,-8)(9,0){4}{\circle{1}}
\multiput(16,-15.5)(9,0){3}{\circle{1}}
\multiput(20.5,-23)(9,0){2}{\circle{1}}
\put(25,-30.5){\circle{1}}
\thicklines
\put(-50,0){\line(3,-5){25}}
\put(-50,0){\line(3,5){50}}
\put(75,-42){\line(-1,0){100}}
\put(75,-42){\line(3,5){25}}
\put(50,84){\line(3,-5){50}}
\put(50,84){\line(-1,0){50}}
\thinlines
\put(-2,-5){\mb0{0}}
\put(15,-48){\mb0{$-\pf$}}
\put(-35,48){\mb0{$\pf$}}
\put(-55,76){\mb0{$\dpf$}}
\put(-55,-8){\mb0{$-\pf$}}
\put(51,-8){\mb0{$\pf$}}
\put(113,-8){\mb0{$\dpf$}}
\end{picture}
\end{center}
\caption{Phase diagram of the honeycomb hard square model.
A, B, C and D are the regions in which the exact solution of the IHS model
is defined.
In order to emphasize the three-fold symmetry of the lattice,
we choose the non-rectangular coordinate system where $v$-axis makes
an angle $120^\circ$ with the $u$-axis.}
\label{phase}
\end{figure}
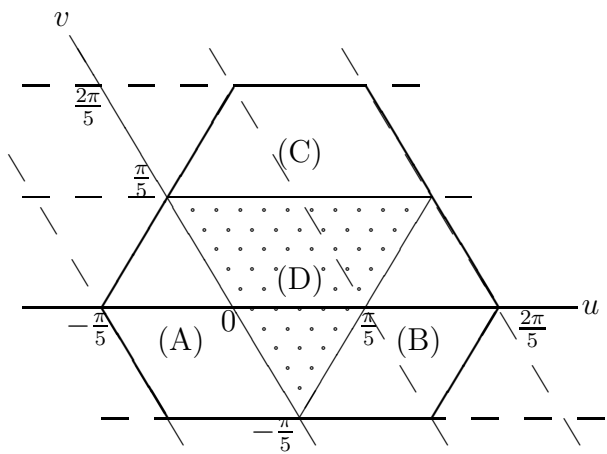

\begin{figure}
\begin{center}
\begin{picture}(300,130)(-7,-30)
\put(40,-20){\mb0{(A)}}\put(140,-20){\mb0{(B)}}
\put(240,-20){\mb0{(C)}}
\multiput(0,0)(20,0){5}{\line(0,-1){7}}
\multiput(0,0)(20,0){4}{\line(5,3){10}}
\multiput(20,0)(20,0){4}{\line(-5,3){10}}
\multiput(10,18)(20,0){4}{\line(0,-1){12}}
\multiput(10,18)(20,0){4}{\line(-5,3){10}}
\multiput(10,18)(20,0){4}{\line(5,3){10}}
\multiput(0,36)(20,0){5}{\line(0,-1){12}}
\multiput(0,36)(20,0){4}{\line(5,3){10}}
\multiput(20,36)(20,0){4}{\line(-5,3){10}}
\multiput(10,54)(20,0){4}{\line(0,-1){12}}
\multiput(10,54)(20,0){4}{\line(-5,3){10}}
\multiput(10,54)(20,0){4}{\line(5,3){10}}
\multiput(0,72)(20,0){5}{\line(0,-1){12}}
\multiput(0,72)(20,0){4}{\line(5,3){10}}
\multiput(20,72)(20,0){4}{\line(-5,3){10}}
\multiput(10,85)(20,0){4}{\line(0,-1){7}}
\multiput(60,0)(-10,18){5}{\circle*{4}}
\put(0,0){\circle*{4}}
\multiput(30,6)(-10,18){4}{\circle*{4}}
\multiput(80,24)(-10,18){4}{\circle*{4}}
\put(80,72){\circle*{4}}

\multiput(100,0)(20,0){5}{\line(0,-1){7}}
\multiput(100,0)(20,0){4}{\line(5,3){10}}
\multiput(120,0)(20,0){4}{\line(-5,3){10}}
\multiput(110,18)(20,0){4}{\line(0,-1){12}}
\multiput(110,18)(20,0){4}{\line(-5,3){10}}
\multiput(110,18)(20,0){4}{\line(5,3){10}}
\multiput(100,36)(20,0){5}{\line(0,-1){12}}
\multiput(100,36)(20,0){4}{\line(5,3){10}}
\multiput(120,36)(20,0){4}{\line(-5,3){10}}
\multiput(110,54)(20,0){4}{\line(0,-1){12}}
\multiput(110,54)(20,0){4}{\line(-5,3){10}}
\multiput(110,54)(20,0){4}{\line(5,3){10}}
\multiput(100,72)(20,0){5}{\line(0,-1){12}}
\multiput(100,72)(20,0){4}{\line(5,3){10}}
\multiput(120,72)(20,0){4}{\line(-5,3){10}}
\multiput(110,85)(20,0){4}{\line(0,-1){7}}
\multiput(110,6)(10,18){5}{\circle*{4}}
\multiput(140,0)(10,18){5}{\circle*{4}}
\multiput(170,6)(10,18){2}{\circle*{4}}
\multiput(100,36)(10,18){3}{\circle*{4}}

\multiput(200,0)(20,0){5}{\line(0,-1){7}}
\multiput(200,0)(20,0){4}{\line(5,3){10}}
\multiput(220,0)(20,0){4}{\line(-5,3){10}}
\multiput(210,18)(20,0){4}{\line(0,-1){12}}
\multiput(210,18)(20,0){4}{\line(-5,3){10}}
\multiput(210,18)(20,0){4}{\line(5,3){10}}
\multiput(200,36)(20,0){5}{\line(0,-1){12}}
\multiput(200,36)(20,0){4}{\line(5,3){10}}
\multiput(220,36)(20,0){4}{\line(-5,3){10}}
\multiput(210,54)(20,0){4}{\line(0,-1){12}}
\multiput(210,54)(20,0){4}{\line(-5,3){10}}
\multiput(210,54)(20,0){4}{\line(5,3){10}}
\multiput(200,72)(20,0){5}{\line(0,-1){12}}
\multiput(200,72)(20,0){4}{\line(5,3){10}}
\multiput(220,72)(20,0){4}{\line(-5,3){10}}
\multiput(210,85)(20,0){4}{\line(0,-1){7}}
\multiput(200,0)(20,0){5}{\circle*{4}}
\multiput(200,24)(20,0){5}{\circle*{4}}
\multiput(210,54)(20,0){4}{\circle*{4}}
\multiput(210,78)(20,0){4}{\circle*{4}}
\end{picture}
\begin{picture}(130,120)(-7,-20)
\multiput(0,0)(20,0){6}{\line(0,-1){7}}
\multiput(0,0)(20,0){5}{\line(5,3){10}}
\multiput(20,0)(20,0){5}{\line(-5,3){10}}
\multiput(10,18)(20,0){5}{\line(0,-1){12}}
\multiput(10,18)(20,0){5}{\line(-5,3){10}}
\multiput(10,18)(20,0){5}{\line(5,3){10}}
\multiput(0,36)(20,0){6}{\line(0,-1){12}}
\multiput(0,36)(20,0){5}{\line(5,3){10}}
\multiput(20,36)(20,0){5}{\line(-5,3){10}}
\multiput(10,54)(20,0){5}{\line(0,-1){12}}
\multiput(10,54)(20,0){5}{\line(-5,3){10}}
\multiput(10,54)(20,0){5}{\line(5,3){10}}
\multiput(0,72)(20,0){6}{\line(0,-1){12}}
\multiput(0,72)(20,0){5}{\line(5,3){10}}
\multiput(20,72)(20,0){5}{\line(-5,3){10}}
\multiput(10,85)(20,0){5}{\line(0,-1){7}}
\multiput(0,0)(20,0){6}{\circle*{4}}
\multiput(10,18)(20,0){5}{\circle*{4}}
\multiput(0,36)(20,0){6}{\circle*{4}}
\multiput(10,54)(20,0){5}{\circle*{4}}
\multiput(0,72)(20,0){6}{\circle*{4}}
\put(60,-16){\mb0{(D)}}
\end{picture}
\end{center}
\caption{The ground states in the interacting
hard square model.
The figure (A), (B), (C) and (D) are the ground states
of region A, B, C and D, of Fig.5, respectively.}
\label{gs}
\end{figure}
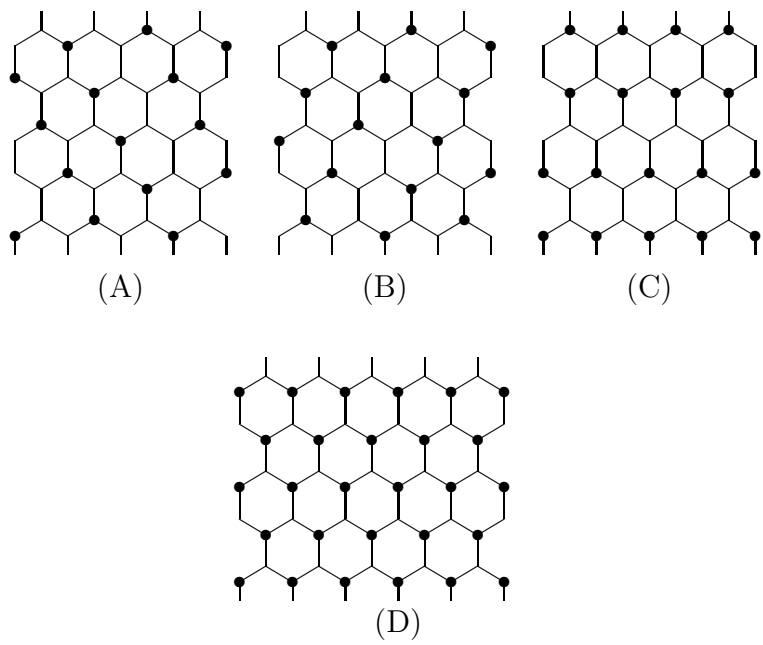

$q^2\rightarrow -1$ is the high temperature limit and $q^2=1$
corresponds to zero temperature.
If we compute the hexagonal face weights for any $u$ and $v$
in these two limits,
the dominant configurations on the honeycomb lattice can be obtained.
By tracing the dominant configurations, we can
draw the phase diagram of honeycomb lattice model.
There are four regions in the $(u$-$v)$ plane where the ground
state configurations take different nature.
This is shown in Fig.\ref{phase}.
The ground state configurations of each region are drawn in Fig.\ref{gs}.
The region D includes the point of $(u,v)=(2\pi/15,\pi/15)$
where the interaction becomes isotropic.
Fig.\ref{gs}(D) shows a ground state of the region D.
Even if we interchange the empty and occupied sites
in the state of Fig.\ref{gs}(D),
it does not affect on the energy of the system.
That is, there are doubly degenerate ground states.
Their particle density is 1/2.
The critical point, corresponding to $q=0$, in this region is
of the tricritical Ising model universality class as in the
original square lattice model.
In the regions A, B, and C, the particle density of the ground state
is 1/3.
The ground states of the region B and C are obtained by rotating
the system of region A by $120^\circ$ and $-120^\circ$,
respectively.
The honeycomb lattice is divided by three sublattices
and the ground states of the region A, B, and C
have three-fold degeneracy.
As the temperature increases from zero, the domain walls
will appear. The system having three different domains
is described by the 3-state Potts model.\cite{domain}
In the high temperature limit of $q^2\rightarrow -1$,
we can see as the dominant phase the vacuum phase in
the regions A, B and C.

Let us call the ground states of (A), (B) and (C) as $3\times 1$ ordering
and that of (D) as $\sqrt 2\times\sqrt 2$ ordering. We borrow this
notation from  the square lattice ordering.
When $v$ approaches to 0,
the system becomes the square lattice and the ground states given in
Fig.\ref{gs} are just the $3\times 1$ and $\sqrt 2\times\sqrt 2$ orderings
on the square lattice.

\begin{figure}
\begin{center}
\begin{picture}(250,230)(-50,-40)
\multiput(0,0)(12,24){7}{\line(1,0){24}}
\multiput(0,0)(12,24){7}{\line(-3,5){12}}
\multiput(0,0)(12,24){7}{\line(-6,-1){24}}
\multiput(24,0)(12,24){7}{\line(6,1){24}}
\multiput(36,24)(12,24){7}{\line(3,-5){12}}
\multiput(24,0)(48,4){2}{\line(3,-5){9}}
\put(-24,136){\line(1,0){24}}
\multiput(-12,68)(12,24){4}{\line(1,0){24}}
\multiput(0,92)(12,24){3}{\line(-3,5){12}}
\multiput(0,92)(12,24){3}{\line(-6,-1){24}}
\put(-28,20){\line(1,0){16}}
\put(-20,44){\line(1,0){20}}
\multiput(48,4)(12,24){7}{\line(1,0){24}}
\multiput(96,8)(12,24){6}{\line(1,0){24}}
\multiput(96,8)(12,24){6}{\line(-3,5){12}}
\multiput(96,8)(12,24){7}{\line(-6,-1){24}}
\multiput(120,8)(12,24){6}{\line(3,-5){12}}
\multiput(120,8)(12,24){6}{\line(6,1){24}}
\multiput(144,12)(12,24){6}{\line(1,0){24}}
\multiput(156,-12)(12,24){6}{\line(6,1){24}}
\multiput(168,12)(12,24){6}{\line(3,-5){12}}
\multiput(180,-8)(12,24){3}{\line(1,0){24}}
\put(216,16){\line(3,-5){10}}
\put(216,16){\line(6,1){18}}
\multiput(216,64)(12,24){2}{\line(1,0){15}}
\multiput(72,144)(36,-20){5}{\circle{30}}
\multiput(0,92)(36,-20){6}{\circle{30}}
\put(0,0){\circle{30}}
\multiput(0,44)(36,-20){3}{\circle{30}}
\multiput(0,136)(36,-20){7}{\circle{30}}
\multiput(144,148)(36,-20){3}{\circle{30}}
\end{picture}
\end{center}
\caption{$3\times 1$ ordering of the region A
of the IHS model represented in the anisotropic lattice system.}
\label{aniso}
\end{figure}
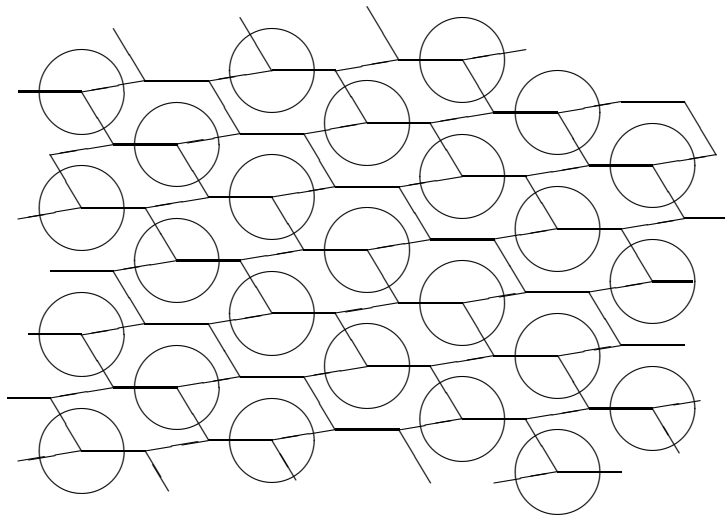

The $3\times 1$ orderings on the honeycomb lattice may look rather strange.
However, by converting the anisotropy of interaction
to the geometrical anisotropy of lattice as discussed in
Ref.\cite{KimPearce}, they are reduced to the compact ordering of
large particles as shown in Fig.\ref{aniso}.
Therefore, the nature of symmetry breaking in the honeycomb lattice
is the same as in the square lattice and no new universality classes appear
in the HLSM.

\section{Magnetic hard square (MHS) model}

The magnetic hard square (MHS) model is the IHS model
where each of particle has spin 1/2 internal degrees of freedom.
When  a site is empty, its spin value is assigned to be zero.
Otherwise, the spin is $1$ or $-1$.
Its face weights are defined by
\bean
\lefteqn{w(a,b,c,d)}& & \\
&=&\left\{\begin{array}{l}
mz^{(a^2+b^2+c^2+d^2)/4}\exp(La^2c^2+Mb^2d^2
+Jac+Kbd)t^{(a^2-b^2+c^2-d^2)/4} \\
\hskip 2in\mbox{, if $ab=bc=cd=da=0$,}\\
0\hskip 1.9in\mbox{, otherwise.}\end{array}
\right.
\eean
The exact solution of this model on the square lattice
which obeys the STR was found by Pearce\cite{MHS}.

For the solution on the so-called E manifold,
the face weights are written by the elliptic theta functions
parametrized by the spectral parameter $u$ and the temperature-like
parameter $q$.
On the E manifold, the spectral parameter varies in
$-\pi/6<u<\pi/2$.
$q$ has the value between $-1$ and $1$.
On the square lattice, $q\rightarrow 1$ is
the zero temperature limit and $q\rightarrow -1$ is
the high temperature limit.
At $q=0$, the E manifold crosses the critical line of T manifold where
all the face weights are the trigonometric functions of $u$.
The T manifold is a line parametrized by the crossing parameter $\lambda$
and the $q=0$ point of the E manifold corresponds to the $\lambda=\pi/3$
point of the T manifold.

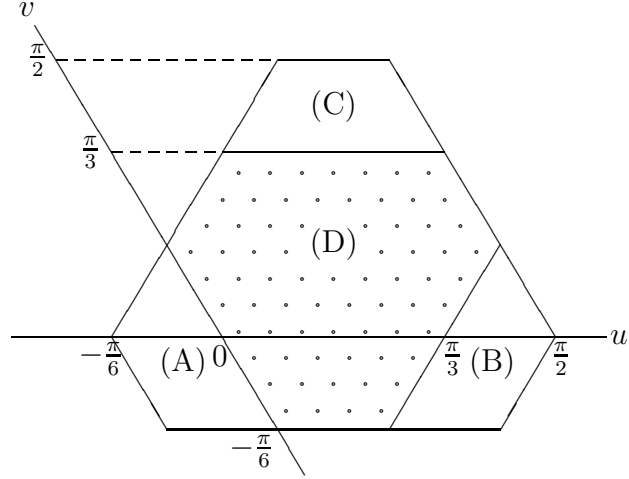
\begin{figure}
\begin{center}
\begin{picture}(230,170)(-85,-55)
\put(31,-52){\line(-3,5){102}}
\put(-80,0){\line(1,0){225}}
\put(150,0){\mb0{$u$}}
\put(-74,124){\mb0{$v$}}
\put(-42,0){\line(3,5){63}}
\put(-42,0){\line(3,-5){21}}
\put(-21,-35){\line(1,0){126}}
\put(126,0){\line(-3,-5){21}}
\put(126,0){\line(-3,5){63}}
\put(21,105){\line(1,0){42}}
\put(0,70){\line(1,0){84}}
\put(63,-35){\line(3,5){42}}
\put(128,-8){\mb0{$\frac{\pi}{2}$}}
\put(11,-43){\mb0{$-\frac{\pi}{6}$}}
\put(-46,-8){\mb0{$-\frac{\pi}{6}$}}
\put(-1,-7){\mb0{$0$}}
\put(-50,70){\mb0{$\frac{\pi}{3}$}}
\put(-70,105){\mb0{$\frac{\pi}{2}$}}
\put(87,-8){\mb0{$\frac{\pi}{3}$}}
\multiput(-42,70)(7,0){6}{\line(1,0){4}}
\multiput(-63,105)(7,0){13}{\line(1,0){4}}
\put(-15,-10){\mb0{(A)}}
\put(102,-10){\mb0{(B)}}
\put(42,35){\mb0{(D)}}
\put(42,87.5){\mb0{(C)}}
\multiput(24,-28)(-6,10){7}{\circle{1}}
\multiput(36,-28)(-6,10){8}{\circle{1}}
\multiput(48,-28)(-6,10){9}{\circle{1}}
\multiput(60,-28)(-6,10){10}{\circle{1}}
\multiput(18,62)(6,-10){3}{\circle{1}}
\multiput(42,22)(6,-10){5}{\circle{1}}
\multiput(30,62)(6,-10){2}{\circle{1}}
\multiput(54,22)(6,-10){4}{\circle{1}}
\multiput(42,62)(6,-10){7}{\circle{1}}
\multiput(54,62)(6,-10){6}{\circle{1}}
\multiput(66,62)(6,-10){5}{\circle{1}}
\multiput(78,62)(6,-10){4}{\circle{1}}
\end{picture}
\end{center}
\caption{Definition of four regions A, B, C and D in the
honeycomb lattice MHS model}
\label{MHSphase}
\end{figure}

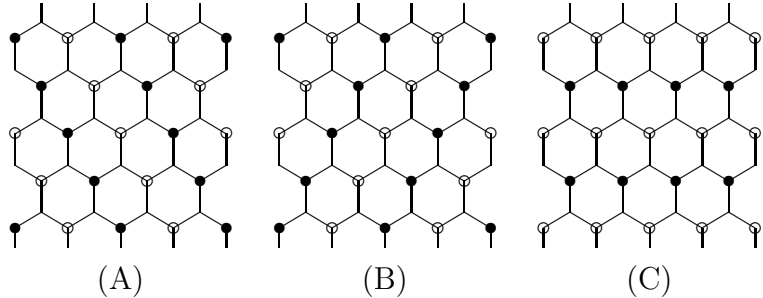
\begin{figure}
\begin{center}
\begin{center}
\begin{picture}(300,130)(-7,-30)
\put(40,-20){\mb0{(A)}}\put(140,-20){\mb0{(B)}}
\put(240,-20){\mb0{(C)}}
\multiput(0,0)(20,0){5}{\line(0,-1){7}}
\multiput(0,0)(20,0){4}{\line(5,3){10}}
\multiput(20,0)(20,0){4}{\line(-5,3){10}}
\multiput(10,18)(20,0){4}{\line(0,-1){12}}
\multiput(10,18)(20,0){4}{\line(-5,3){10}}
\multiput(10,18)(20,0){4}{\line(5,3){10}}
\multiput(0,36)(20,0){5}{\line(0,-1){12}}
\multiput(0,36)(20,0){4}{\line(5,3){10}}
\multiput(20,36)(20,0){4}{\line(-5,3){10}}
\multiput(10,54)(20,0){4}{\line(0,-1){12}}
\multiput(10,54)(20,0){4}{\line(-5,3){10}}
\multiput(10,54)(20,0){4}{\line(5,3){10}}
\multiput(0,72)(20,0){5}{\line(0,-1){12}}
\multiput(0,72)(20,0){4}{\line(5,3){10}}
\multiput(20,72)(20,0){4}{\line(-5,3){10}}
\multiput(10,85)(20,0){4}{\line(0,-1){7}}
\put(0,0){\circle*{4}}
\multiput(20,0)(-10,18){3}{\circle{4}}
\multiput(40,0)(-10,18){5}{\circle*{4}}
\multiput(60,0)(-10,18){5}{\circle{4}}
\multiput(80,0)(-10,18){5}{\circle*{4}}
\multiput(80,36)(-10,18){3}{\circle{4}}
\put(80,72){\circle*{4}}

\multiput(100,0)(20,0){5}{\line(0,-1){7}}
\multiput(100,0)(20,0){4}{\line(5,3){10}}
\multiput(120,0)(20,0){4}{\line(-5,3){10}}
\multiput(110,18)(20,0){4}{\line(0,-1){12}}
\multiput(110,18)(20,0){4}{\line(-5,3){10}}
\multiput(110,18)(20,0){4}{\line(5,3){10}}
\multiput(100,36)(20,0){5}{\line(0,-1){12}}
\multiput(100,36)(20,0){4}{\line(5,3){10}}
\multiput(120,36)(20,0){4}{\line(-5,3){10}}
\multiput(110,54)(20,0){4}{\line(0,-1){12}}
\multiput(110,54)(20,0){4}{\line(-5,3){10}}
\multiput(110,54)(20,0){4}{\line(5,3){10}}
\multiput(100,72)(20,0){5}{\line(0,-1){12}}
\multiput(100,72)(20,0){4}{\line(5,3){10}}
\multiput(120,72)(20,0){4}{\line(-5,3){10}}
\multiput(110,85)(20,0){4}{\line(0,-1){7}}
\multiput(100,0)(10,18){5}{\circle*{4}}
\multiput(120,0)(10,18){5}{\circle{4}}
\multiput(140,0)(10,18){5}{\circle*{4}}
\multiput(160,0)(10,18){3}{\circle{4}}
\put(180,0){\circle*{4}}
\multiput(100,36)(10,18){3}{\circle{4}}
\put(100,72){\circle*{4}}

\multiput(200,0)(20,0){5}{\line(0,-1){7}}
\multiput(200,0)(20,0){4}{\line(5,3){10}}
\multiput(220,0)(20,0){4}{\line(-5,3){10}}
\multiput(210,18)(20,0){4}{\line(0,-1){12}}
\multiput(210,18)(20,0){4}{\line(-5,3){10}}
\multiput(210,18)(20,0){4}{\line(5,3){10}}
\multiput(200,36)(20,0){5}{\line(0,-1){12}}
\multiput(200,36)(20,0){4}{\line(5,3){10}}
\multiput(220,36)(20,0){4}{\line(-5,3){10}}
\multiput(210,54)(20,0){4}{\line(0,-1){12}}
\multiput(210,54)(20,0){4}{\line(-5,3){10}}
\multiput(210,54)(20,0){4}{\line(5,3){10}}
\multiput(200,72)(20,0){5}{\line(0,-1){12}}
\multiput(200,72)(20,0){4}{\line(5,3){10}}
\multiput(220,72)(20,0){4}{\line(-5,3){10}}
\multiput(210,85)(20,0){4}{\line(0,-1){7}}
\multiput(200,0)(20,0){5}{\circle{4}}
\multiput(210,18)(20,0){4}{\circle*{4}}
\multiput(200,36)(20,0){5}{\circle{4}}
\multiput(210,54)(20,0){4}{\circle*{4}}
\multiput(200,72)(20,0){5}{\circle{4}}
\end{picture}
\end{center}
\end{center}
\caption{ Antiferromagnetic orderings which
are the ground states of the region A, B and C, respectively.}
\label{AF}
\end{figure}

Applying the elliptic solution of the square lattice
to the honeycomb lattice model,
we can draw the phase diagram of the honeycomb lattice
MHS model.
It is divided into four regions as shown in Fig.\ref{MHSphase}.
At zero temperature, the particles form
$\sqrt 2\times\sqrt 2$ ordering as in the previous section.
In the $\sqrt 2\times\sqrt 2$ ordering the honeycomb lattice has
two sublattices.
The ground states are that the sites of one sublattice are occupied
and those of the other are empty.
In the region D including the isotropic point,
the system shows the ferromagnetic phase at zero temperature limit.
In the other regions the antiferromagnetic
phases are dominant
where $+1$ and $-1$ spins are alternating line by line.
In the square lattice limit, this corresponds to
the $(01\bar 01)$ state, following the notation used in
Ref.\cite{Kuniba}.
Fig.\ref{AF} illustrates the antiferromagnetic phases
of the MHS model.
In the high temperature limit, the region D
is the paramagnetic phase where the spins
of the occupied sublattice are
distributed randomly and the local magnetization is zero.
This limit belongs to the non-magnetic IHS model
discussed in the previous section.
Since there is no magnetic interaction in the IHS model,
neither spin is preferred and the system shows
the random distribution of the spin.
In the other regions, the vacuum  state is  dominant.

The T manifold is the critical line where the face weights are
the trigonometric functions of the spectral parameter.
On the T manifold, the bulk free energy is given by  that
of the six-vertex model.
This manifold is classified according to the spectral
parameter $u$ into
three regions called as I/II, III/IV and V/VI
region.\footnote{The word {\it regime} is conventionally used
instead {\it region} for classifying the regions according to
the spectral parameter. But the word {\it region} would be
more proper to imply the range depending on the spectral parameter.
In this paper, we choose the word {\it region}.}
In I/II region, $u$ varies in $-(\pi-\lambda)<u<0$ and
in III/IV region, in $0<u<\lambda$.
V/VI region is where $u$ is in the range $\lambda<u<\pi$.
In each region, the anisotropy angle
of the square lattice
which appears in the finite-size scaling is
found\cite{KimPearce,I/II,thesis} to be
\beq
\theta=\left\{
\begin{array}{lll}
-u/(1-\lambda/\pi)&\mbox{\ \  }&\mbox{in region I/II,}\\
u\pi/\lambda&\mbox{\ }&\mbox{in region III/IV,}\\
(\pi-u)/(1-\lambda/\pi)&\mbox{\ }&\mbox{in region V/VI.}
\end{array}\right.
\eeq
The central charge of the square lattice model is $c=1$\cite{III/IV}.

By calculating numerically the eigenvalues of the transfer matrix
for the honeycomb lattice model and using the finite-size scaling
proposed in Sec.II, the central charge and the scaling dimensions are obtained.
{}From the numerical results, this model is manifested
to have the central charge $c=1$.
In the case of $\lambda=5\pi/12$, $u=\lambda/3$ and
$v=\lambda/8$, we calculated the eigenvalues for $N=4$, $5$ and
$6$. It is the case where all of three faces constituting one
hexagonal face are in III/IV region.
\begin{table}
\begin{center}
\begin{tabular}{c|c|c|cc}
\hline
[$S$,$\tilde r$] & $N=6$ & Extrapolated & \multicolumn{2}{c}
{Exact value}\\ \hline
[$0,+$] & $-1.6\times 10^{-4}$ &
$1.4\times 10^{-5}$ & $0$& $(0,0)$ \\
& $0.0953$ &$0.0952$ & $0.0952$ & $X_{1,0}$ \\
& $0.3803$ & $0.3810$ & $0.3810$ & $X_{2,0}$ \\
& $0.8725$ & $0.8543$ & $0.857$ & $X_{3,0}$ \\
& $1.501$ & $1.523$ & $1.524$ & $X_{4,0}$ \\
& $1.965$ & $2.000$ & $2$ & $(1,1)$ \\
& $2.040$ & $2.099$ & $2.095$ & $X_{1,0}+1+1$ \\
& $2.317$ & $2.374$ & $2.381$ & $X_{5,0}$ \\
& $2.342$ & $2.373$ & $2.381$ & $X_{2,0}+1+1$ \\
& $2.562$ & $2.620$ & $2.857$ & $X_{3,0}+1+1$ \\ \hline
[$0,-$] & *$0.1254$ & $0.1251$ & $0.125$ & $(1/16,1/16)$ \\
& *$1.127$ & $1.128$ & $1.125$ & $(9/16,9/16)$  \\
& *$2.121$ & $2.131$ & $2.125$ & $(1/16+1, 1/16+1)$ \\ \hline
[$1,+$] & $1.089$ & $1.094$ & $1.095$ & $X_{1,0}+1$ \\
& $1.362$ & $1.377$ & $1.381$ & $X_{2,0}+1$ \\
& $1.861$ & $1.870$ & $1.857$ & $X_{3,0}+1$ \\
& $2.421$ & $2.503$ & $2.524$ & $X_{4,0}+1$ \\
& $2.652$ & $2.706$ & -- \\
& $2.887$ & $2.955$ & -- \\ \hline
[$1,-$]& *$1.120$ & $1.125$ & $1.125$ & $(1/16+1,1/16)$ \\
& *$2.069$ & $2.099$ & $2.125$ & $(9/16+1,9/16)$ \\
& *$2.092$ & $2.131$ & $2.125$ & $(25/16,9/16)$ \\ \hline
[$2,+$] & $1.962$ & --
& $2$ & $(0+2,0)$ \\
& $2.010$ & $2.078$ & $2.095$ & $X_{1,0}+2$ \\
& $2.065$ & -- & $2.095$ & $X_{1,0}+2$ \\
& $2.222$ & $2.624$ & $2.381$ & $X_{2,0}+2$ \\
& $2.343$ & -- & $2.381$ & $X_{2,0}+2$ \\ \hline
[$2,-$] & *$2.018$ & $2.112$ & $2.125$ & $(1/16+2,1/16)$ \\
& *$2.080$ & $2.112$ & $2.125$ & $(1/16+2,1/16)$ \\ \hline
\end{tabular}
\end{center}
\caption{Transfer matrix spectrum of the honeycomb lattice
MHS model with $\lambda=5\pi/12$, $u=\lambda/3$ and
$v=\lambda/8$. The extrapolation is done from the data
of $N=4$, $5$, and $6$. The starred levels are doublets.}
\label{tab1}
\end{table}
Table \ref{tab1} shows the data of $N=6$ for the scaling dimensions
and the extrapolated values.
We extrapolate the results by fitting finite size data to the form
$x_\alpha(N)=x_\alpha(\infty)+a/N^b$ with
a constant $a$ and $b$ to get the $N$-independent scaling dimension
$x_\alpha(\infty)$.
We divide the levels into sectors denoted as $[S,\tilde r]$.
$S$ is the spin of the field and $\tilde r$ is the eigenvalue
for the spin reversal operator.
Their rigorous definition is
given in Ref.\cite{I/II}.
The starred levels in the table represent doublets.
For other notations, see Ref.\cite{III/IV} and \cite{I/II}.
The transfer matrix spectrum shows that the honeycomb lattice
MHS model belongs to Ashkin-Teller
class as in the case of the square lattice model.
The last column of the table provides the corresponding
scaling dimensions of the Ashkin-Teller model.
$(0,0)$ is the level related to the identity operator.
$X_{n,m}$'s are the scaling dimensions of the gaussian operators,
which are given by
\[X_{n,m}=\frac{n^2}{2g}+\frac{g}{2}m^2\;.\]
where $n$, $m$ are integers and $g$ is the gaussian model
coupling constant and is related to the crossing parameter by
$g=9(1-\lambda/\pi)$.
The dimensions (1/16, 1/16), (9/16, 9/16) and (25/16, 9/16)
are associated with the magnetization operator and are $\lambda$-independent.

When all the values of $u$, $v$ and $(u-v)$ are in the region I/II
or V/VI, we also have the same result as above.
But in the cases that the faces of the region III/IV are mixed
with those of the region I/II or V/VI, it appears that there
exist an additional factor in the bulk free energy term
besides the sum of the bulk free energy of each face.
Since we could not calculate the spectra for large enough system,
we could not resolve the nature of this anomaly.
Resumably, it comes from the mismatch between the
ground states of different regions.
Even if there exist an unexplained additional term, the $1/N$ correction
term in the finite-size scaling show the  expected behavior
of Ashkin-Teller class for all the cases
we have calculated.

The MHS model is generalized to the affine D model.
The affine D model is characterized by $\lambda$ and an integer $L$.
The number of spin states is $(L+3)$.
The MHS model is equivalent to the affine D model with $L=3$.
In the general affine D model, the exact solution
is found by Kuniba and Yajima\cite{Kuniba} when $\lambda=\pi/L$.
(Pasquier's ADE model corresponds to $\lambda=0$.\cite{ADE})
The transfer matrix spectra of the square lattice affine D model
are analyzed in Ref.\cite{affine,I/II}.
It is found that they belong to the Ashkin-Teller universality class
with the gaussian coupling given by $g=L^2(1-\lambda/\pi)$.
For the general affine D model, the honeycomb lattice model
is also defined.
We have also analyzed the transfer matrix spectrum
for several cases of the honeycomb lattice affine D model and have found
that their critical properties are again
the same as in the square lattice model.

\section{Extension of the honeycomb lattice solvable models}

Up to now we have investigated some properties of
the honeycomb lattice models where each hexagonal face weights are given
by the multiplication of the square face weights with the parameter
$u$, $(u-v)$ and $v$.
In this section we extend this scheme of constructing the integrable
models to another kind of integrable models with larger unit cells.
A hexagonal face consists of three parallelograms and
we will distinguish the spectral parameters for
the upper-right, upper-left and lower parallelograms by
$u$, $u^\prime$ and $u^{''}$, respectively.
For the models defined in Sec.II, they are fixed to be $u^\prime=u-v$
and $u^{''}=v$.
When the size of a unit cell is given as
$m\times n$ with positive integers $m$ and $n$,
let us label the spectral parameters of the hexagon
located at $i$-th column and $j$-th
row of a unit cell as $u_{ij}$, $u^\prime_{ij}$ and $u^{''}_{ij}$.
Here $i$($j$) is the positive integer modulo $m$($n$).
\begin{figure}
\begin{center}
\begin{picture}(370,200)(-50,0)
\multiput(40,190)(60,0){4}{\line(-5,-3){30}}
\multiput(40,190)(60,0){4}{\line(5,-3){30}}
\multiput(10,172)(60,0){5}{\line(0,-1){36}}
\multiput(40,118)(60,0){4}{\line(5,3){30}}
\multiput(40,118)(60,0){5}{\line(-5,3){30}}
\multiput(40,118)(60,0){5}{\line(0,-1){36}}
\multiput(40,154)(60,0){4}{\line(0,1){36}}
\multiput(40,154)(60,0){4}{\line(5,-3){30}}
\multiput(40,154)(60,0){4}{\line(-5,-3){30}}
\multiput(70,64)(60,0){4}{\line(5,3){30}}
\multiput(70,64)(60,0){5}{\line(-5,3){30}}
\multiput(70,64)(60,0){5}{\line(0,-1){36}}
\multiput(70,100)(60,0){4}{\line(0,1){36}}
\multiput(70,100)(60,0){4}{\line(5,-3){30}}
\multiput(70,100)(60,0){4}{\line(-5,-3){30}}
\multiput(100,10)(60,0){4}{\line(5,3){30}}
\multiput(100,10)(60,0){4}{\line(-5,3){30}}
\multiput(100,46)(60,0){4}{\line(0,1){36}}
\multiput(100,46)(60,0){4}{\line(5,-3){30}}
\multiput(100,46)(60,0){4}{\line(-5,-3){30}}
\put(25,167){\mb0{$u_{11}^{'}$}}\put(55,167){\mb0{$u_{11}$}}
\put(40,136){\mb0{$u_{11}^{''}$}}
\put(85,167){\mb0{$u_{12}^{'}$}}\put(115,167){\mb0{$u_{12}$}}
\put(100,136){\mb0{$u_{12}^{''}$}}
\put(145,167){\mb0{$u_{13}^{'}$}}\put(175,167){\mb0{$u_{13}$}}
\put(160,136){\mb0{$u_{13}^{''}$}}
\put(205,167){\mb0{$u_{14}^{'}$}}\put(235,167){\mb0{$u_{14}$}}
\put(220,136){\mb0{$u_{14}^{''}$}}
\put(55,113){\mb0{$u_{21}^{'}$}}\put(85,113){\mb0{$u_{21}$}}
\put(70,82){\mb0{$u_{21}^{''}$}}
\put(115,113){\mb0{$u_{22}^{'}$}}\put(145,113){\mb0{$u_{22}$}}
\put(130,82){\mb0{$u_{22}^{''}$}}
\put(175,113){\mb0{$u_{23}^{'}$}}\put(205,113){\mb0{$u_{23}$}}
\put(190,82){\mb0{$u_{23}^{''}$}}
\put(235,113){\mb0{$u_{24}^{'}$}}\put(265,113){\mb0{$u_{24}$}}
\put(250,82){\mb0{$u_{24}^{''}$}}
\put(85,59){\mb0{$u_{31}^{'}$}}\put(115,59){\mb0{$u_{31}$}}
\put(100,28){\mb0{$u_{31}^{''}$}}
\put(145,59){\mb0{$u_{32}^{'}$}}\put(175,59){\mb0{$u_{32}$}}
\put(160,28){\mb0{$u_{32}^{''}$}}
\put(205,59){\mb0{$u_{33}^{'}$}}\put(235,59){\mb0{$u_{33}$}}
\put(220,28){\mb0{$u_{33}^{''}$}}
\put(265,59){\mb0{$u_{34}^{'}$}}\put(295,59){\mb0{$u_{34}$}}
\put(280,28){\mb0{$u_{34}^{''}$}}

\end{picture}
\end{center}
\caption{A unit cell consisting of $3\times 4$ hexagons}
\label{fusion}
\end{figure}
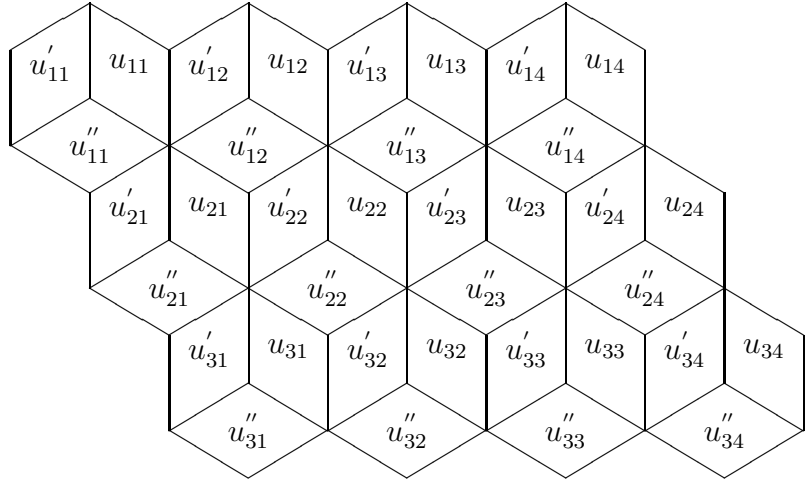
In Fig.\ref{fusion}, the unit cell of $3\times 4$ hexagons
is illustrated as an example.
Suppose that the spectral parameters of parallelograms
included in the unit cell are related by
two variables of ($u$, $v$) as following:
\bea
u_{ij}&=&u+(i-1)\cdot(u-v) \nonumber \\
u^\prime_{ij}&=&u_{ij}-(n-j+1)v \label{uij} \\
u^{''}_{ij}&=&(n-j+1)v\; .\nonumber
\eea
As long as the spectral parameters satisfy Eq.(\ref{uij}),
the integrability of this model also can be proved by the
STR in the simple way.
Thus one can generate a hierarchy of integrable models starting from each
integrable square lattice model.

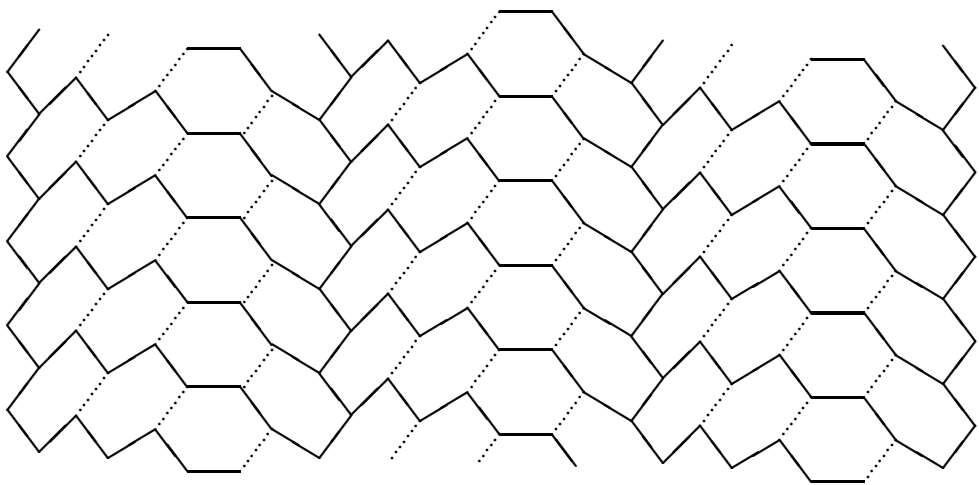
\begin{figure}
\begin{center}
\begin{picture}(380,220)(-20,-50)
\dotl{27.5,0}\dotl{27.5,32}\dotl{27.5,64}\dotl{27.5,96}\dotl{27.5,128}
\dotl{57.5,-5.2}\dotl{57.5,26.8}\dotl{57.5,58.8}
\dotl{57.5,90.8}\dotl{57.5,122.8}
\dotl{89.5,-21.2}\dotl{89.5,8.8}\dotl{89.5,40.8}
\dotl{89.5,72.8}\dotl{89.5,104.8}
\dotl{145.5,-18}\dotl{145.5,14}
\dotl{145.5,46}\dotl{145.5,78}\dotl{145.5,110}
\dotl{175.5,8.8}\dotl{175.5,40.8}
\dotl{175.5,72.8}\dotl{175.5,104.8}\dotl{175.5,136.8}
\dotl{207.5,-7.2}\dotl{207.5,24.8}
\dotl{207.5,56.8}\dotl{207.5,88.8}\dotl{207.5,120.8}
\multiput(178.5,-19.2)(1.5,2){6}{\circle*{0.1}}
\dotl{263.5,-4}\dotl{263.5,28}\dotl{263.5,60}\dotl{263.5,92}\dotl{263.5,124}
\dotl{293.5,-9.2}\dotl{293.5,22.8}
\dotl{293.5,54.8}\dotl{293.5,86.8}\dotl{293.5,118.8}
\dotl{325.5,-25.2}\dotl{325.5,6.8}
\dotl{325.5,38.8}\dotl{325.5,70.8}\dotl{325.5,102.8}

\thicklines
\multiput(0,0)(0,32){5}{\line(3,-4){12}}
\multiput(0,0)(0,32){5}{\line(3,4){12}}
\multiput(12,-16)(0,32){5}{\line(1,1){14}}
\multiput(26,-2)(0,32){5}{\line(3,-4){12}}
\multiput(38,-18)(0,32){5}{\line(5,3){18}}
\multiput(56,-7.2)(0,32){5}{\line(3,-4){12}}
\multiput(68,-23.2)(0,32){6}{\line(1,0){20}}
\multiput(88,8.8)(0,32){5}{\line(3,-4){12}}
\multiput(100,-7.2)(0,32){5}{\line(5,-3){18}}
\multiput(118,14)(0,32){5}{\line(3,-4){12}}
\multiput(118,-18)(0,32){5}{\line(3,4){12}}
\multiput(130,-2)(0,32){5}{\line(1,1){14}}
\multiput(144,12)(0,32){5}{\line(3,-4){12}}
\multiput(156,-4)(0,32){5}{\line(5,3){18}}
\multiput(174,6.8)(0,32){5}{\line(3,-4){12}}
\multiput(186,-9.2)(0,32){6}{\line(1,0){20}}
\multiput(206,22.8)(0,32){5}{\line(3,-4){12}}
\put(206,-9.2){\line(3,-4){9}}
\multiput(218,6.8)(0,32){5}{\line(5,-3){18}}
\multiput(236,-4)(0,32){5}{\line(3,4){12}}
\multiput(236,-4)(0,32){5}{\line(3,-4){12}}
\multiput(248,-20)(0,32){5}{\line(1,1){14}}
\multiput(262,-6)(0,32){5}{\line(3,-4){12}}
\multiput(274,-22)(0,32){5}{\line(5,3){18}}
\multiput(292,-11.2)(0,32){5}{\line(3,-4){12}}
\multiput(304,-27.2)(0,32){6}{\line(1,0){20}}
\multiput(324,4.8)(0,32){5}{\line(3,-4){12}}
\multiput(336,-11.2)(0,32){5}{\line(5,-3){18}}
\multiput(354,-22)(0,32){5}{\line(3,4){12}}
\multiput(354,10)(0,32){5}{\line(3,-4){12}}
\thinlines
\end{picture}
\end{center}
\caption{ An example of the integrable model with
$1\times 4$ unit cell where the anisotropic interaction is interpreted
as the geometrical anisotropy.
The thick line denotes boundaries of unit cells.}
\label{deform}
\end{figure}

As the honeycomb lattice solvable model corresponds to
the deformed honeycomb lattice as shown in Fig.\ref{aniso},
the system whose unit cell is made of $m\times n$ hexagons with different
anisotropy angles results in the periodic deformation of
honeycomb lattice. One example is illustrated in Fig.\ref{deform}
for the case $m=1$ and $n=4$.

\section{Discussion}
In this paper, we generalized the honeycomb lattice Ising model
which is the IRF version of the kagome lattice eight-vertex model
and have obtained the general honeycomb lattice solvable models.
These models includes two spectral parameters which control deformations
of a square into a parallelogram of general shape.
The finite-size scaling behavior of
the free energy and the anisotropy factor
are written in terms of the quantities obtained
in the square lattice models.
{}From the several solvable models satisfying the STR,
we analyzed the corresponding honeycomb lattice models.
As to the critical phenomena, we do not expect new universality classes
appear in the models constructed in this way,
because the two point correlation function of the honeycomb lattice
model behaves as in the regular square lattice model.
However, some noble behaviors may appear in the non-universal
quantities if the system has different properties in the different
regions of the spectral parameter.

The IRF version of the Z-invariant model is
described by the planar lattice system where
the parallelograms of arbitrary angles fill the plane
with no overlap and no gap.
Considering a hexagon in such a lattice system,
we can change the positions of three parallelograms
constituting the hexagon by applying the STR.
This process does not change the partition function.
It allows us to construct lots of different systems
having the same partition function, that is, the same free energy.
If we consider the IRF version of the kagome lattice model,
there are three kinds of the parallelograms.
Keeping the free energy constant, we can change the distribution
of the parallelograms by using the STR, repeatedly.
Configurations obtained in this way will be the configurations
of the random tiling model discussed in Ref.\cite{tiling}.
The Boltzmann weights of solvable models are analogous to the
chemical potentials of the tiles.
It also can be interpreted as the state of the zero-temperature
triangular Ising antiferromagnet\cite{TIA}, if we use only three
types of parallelograms in the solvable models.
Possible application of the STR to the random tiling problems is left
for future study.
\vskip 0.4in
\noindent{\Large\bf Acknowledgement}
\vskip 0.2in

This work is supported by KOSEF through the Center for Theoretical
Physics, Seoul National University.


\begin{thebibliography}{99}
\bibitem{Baxter} R.J. Baxter, {\it Exactly Solved Models in
Statistical Mechanics}
(Academic Press, New York,1982)
\bibitem{Royal} R.J. Baxter, {\it Phil. Trans. of the
Royal Soc. of London A}{\bf 289}, 315 (1978)
\bibitem{KimPearce}D. Kim and P.A. Pearce, {\it J. Phys. A:Math. Gen.}
{\bf 20}, L451 (1987)
\bibitem{ihs} R.J. Baxter, {\it J. Phys. A:Math. Gen.}
{\bf 13}, L61 (1980)
\bibitem{domain} M. den Nijs, in {\it
Phase Transition and Critical Phenomena},
edited by C. Domb and J.L. Lebowitz, vol.12 (Academic Press, New York,
1988)
\bibitem{MHS} P.A. Pearce, {\it J. Phys. A: Math. Gen.}
{\bf 18}, 3217 (1985)
\bibitem{III/IV} D. Kim, P.A. Pearce, {\it J. Phys. A: Math. Gen.},
{\bf 20}, 6471 (1987); D. Kim, J.-Y. Choi, and K.-H. Kwon,
{\it J. Phys. A: Math. Gen.} {\bf 21}, 2661 (1988)
\bibitem{I/II} K.-H. Kwon, D. Kim, {\it J. of Korean Phys. Soc.}
{\bf 25}, 1 (1992)
\bibitem{thesis} K.-H. Kwon, Ph.D thesis (Seoul National University, 1993)
\bibitem{ADE} V. Pasquier, {\it Nucl. Phys. B}{\bf 285} [{\bf
FS19}], 162 (1987)
\bibitem{Kuniba} A. Kuniba, T. Yajima, {\it J. Stat. Phys.} {\bf 52},
829 (1987)
\bibitem{affine} J.-Y. Choi, K.-H. Kwon and D. Kim, {\it Europhys. Lett.}
{\bf 10}(8), 703 (1989)
\bibitem{tiling} W. Li, H. Park and M. Widom, {\it J. Phys. A: Math. Gen.}
{\bf 23}, L573 (1990)
\bibitem{TIA} H.W.J. Bl\"{o}te, H.J. Hilhorst, {\it J. Phys. A: Math. Gen.}
{\bf 15}, L631 (1982)
\end{thebibliography}
\end{document}